# Optimize Individualized Energy Delivery for Septic Patients Using Predictive Deep Learning Models: A Real World Study

**Running title:** To optimize energy delivery for sepsis in ICU.

Lu Wang, MCs[1,2], Li Chang, BS[3], Ruipeng Zhang, BS[1,2], Kexun Li, MCs[4], Yu Wang, MCs[5], Wei Chen, PhD[5], Xuanlin Feng, BS[3], Mingwei Sun, BS[1,2], Qi Wang, PhD[6], Charles Damien Lu, PhD[2], Jun Zeng, BS[1,2], Hua Jiasng, PhD[1,2,3]

[1] Institute for Emergency and Disaster Medicine, Sichuan Provincial People's Hospital, School of Medicine, University of Electronic Science and Technology of China, Chengdu, 610072, China
[2] Sichuan Provincial Research Center for Emergency Medicine and Critical Illness, Sichuan Provincial People's Hospital, School of Medicine, University of Electronic Science and Technology of China, Chengdu, 610072, China
[3] Department of Emergency Intensive Care Unit, Sichuan Provincial People's Hospital, School of Medicine, University of Electronic Science and Technology of China, Chengdu, 610072, China
[4] Department of Emergency Surgery, Sichuan Provincial People's Hospital, School of Medicine, University of Electronic Science and Technology of China, Chengdu, 610072, China
[5] Department of Clinical Nutrition, Department of Health Medicine, Peking Union Medical College Hospital, Chinese Academy of Medical Sciences and Peking Union Medical College, Beijing, 100730, China
[6] Department of Mathematics, University of South Carolina, Columbia, SC, 29208, USA

Authors' email addresses and contributions:
LW:
Contribution: conceived and delineated the hypotheses, designed the study, acquired, data collection, data synthesis, model building and analyzed the data, and drafted the first version of this manuscript and edited the manuscript of the previous analysis.

LC:
Contribution: developed the original concepts for this study.

R-PZ:
Contribution: data collection, data synthesis and model building.

K-XL:
Contribution: data collection, data synthesis and model building.

YW:
Contribution: drafted the first version of this manuscript.

WC:
Contribution: developed the original concepts for this study.

X-LF:
Contribution: data collection, data synthesis and model building.

M-WS:
Contribution: developed the original concepts for this study.

QW:
Contribution: the English proofreading and editing of this manuscript.

Dr. Charles Damien Lu:
Contribution: the English proofreading and editing of this manuscript.

Lu Wang and Li Chang equally contributed to this work.

**Corresponding Author:** Dr. Hua Jiang, Institute for Emergency and Disaster Medicine, Sichuan Provincial People's Hospital, School of Medicine, University of Electronic Science and Technology of China, Chengdu, 610072, China. Email: jianghua@uestc.edu.cn.

Dr. Jun Zeng, Sichuan Provincial Research Center for Emergency Medicine and Critical Illness, Sichuan Provincial People's Hospital, School of Medicine, University of Electronic Science and Technology of China, Chengdu, 610072, China. Email: zengjun@med.uestc.edu.cn.

ABSTRACT
**Background and Objectives:** We aim to establish deep learning models to optimize the individualized energy delivery for septic patients.
**Methods and Study Design:** We conducted a study of adult septic patients in Intensive Care Unit (ICU), collecting 47 indicators for 14 days. After data cleaning and preprocessing, we used stats to explore energy delivery in deceased and surviving patients. We filtered out nutrition-related features and divided the data into three metabolic phases: acute early, acute late, and rehabilitation. Models were built using data before September 2020 and validated on the rest. We then established optimal energy target models for each phase using deep learning.

**Results:** A total of 277 patients and 3115 data were included in this study. The models indicated that the optimal energy targets in the three phases were 900kcal/d, 2300kcal/d, and 2000kcal/d, respectively. Excessive energy intake increased mortality rapidly in the early period of the acute phase. Insufficient energy in the late period of the acute phase significantly raised the mortality of septic patients. For the rehabilitation phase, too much or too little energy delivery both associated with high mortality.

**Conclusion:** Our study established time-series prediction models for septic patients to optimize energy delivery in the ICU. This approach indicated the feasibility of developing nutritional tools for critically ill patients. We recommended permissive underfeeding only in the early acute phase. Later, increased energy intake may improve survival and settle energy debts caused by underfeeding.

**INTRODUCTION**

Septic patients in intensive care unit (ICU) often suffer from malnutrition. As a result, nutrition support has become one of the most essential therapies. [1,2,3] However, optimal energy targets for critically ill patients in ICU admission remain controversial. [4,5] Energy expenditure (EE) was recommended to guide nutrition in critically ill patients, and the measurements included direct calorimetry, indirect calorimetry, and predictive formula. Indirect calorimetry was the gold standard for measuring EE, but it was difficult to put into clinical practice due to its expensive equipment and complicated operation. [6] Nowadays, clinical practitioners prefer to apply prediction formulas because of their simplicity and convenience. However, recent studies indicated that they did not work as well as perceived by clinical practitioners. [7,8] An applicable and easy-to-use tool is urgently in need that can help clinicians to predict the optimal energy target for septic patients. Although there have been many attempts, most did not work well when implemented in clinical settings. From a data science perspective, the major reason for the failure is due to the "curse of dimensionality", [9] which is a phenomenon in vector problems where the amount of computation increases exponentially as the number of dimensions increases. In the ICU setting, deciding the nutritional therapy protocol for a specific patient require the consideration of many influencing factors represented by a range of clinical indicators (e.g., dozens of biochemical and blood test parameters, patients' nutritional data, etc.). With the rapid development of data science, deep learning (DL) has shown as a powerful tool to take on such challenges since researchers have realized that applying DL may help to reduce dimensions and thus overcome the curse of dimensionality. [10,11]

In 2011, the tight calorie control study (TICACOS) indicated that both overfeeding and underfeeding were associated with a higher risk of death in critically ill patients.[12]

Recently, several researchers introduced the concept of identifying the "sweet spot" (the spot where nutrition may have a positive impact on outcomes) or "personalized target" for nutrition doses based on the severity of illness/nutrition risks.[13] In this study, we propose the hypothesis that for septic patients in different metabolic phases, there is an individual optimal energy target or the sweet spot that can minimize the risk of death. Subsequently, we aim to establish deep learning models to determine such energy targets.

MATERIALS AND METHODS

*The aim*

We aim to establish deep learning models to optimize the individualized energy delivery for septic patients.

*Include Criteria and Exclusion Criteria*

Our study enrolled hospital patients suffering from sepsis in an intensive care unit (ICU) of a single center between September 2018 and January 2023. The Inclusion criteria are as follows:

① Age ≥ 18 years (adult patients);
② Patients who meet the criteria of Sepsis 3.0;
③ APACHE II score ≥ 10;

And the exclusion criteria are as follows:

① Age <18 years old;
② Patients who used extracorporeal membrane oxygenation (ECMO) or renal replacement therapy in the treatment;
③ Women who are pregnant or breastfeeding;
④ Patients participating in other clinical trials.

Our study has been reviewed by the Medical Ethics Committee of Sichuan Provincial People's Hospital (No. 190 in 2019) and has been registered in the China Clinical Trial Registration Center (Clinical Registration No.: ChiCTR1900024746. Registered 26 July 2019. https://www.chictr.org.cn/index.html). This study is an observational study and will not interfere with the treatment plan of patients, the ethics committee waived informed consent.

*Data Collection and cleaning*

The collected indicators included both static and dynamic indicators. The static indicators were cross-sectional data at the time of patient admission and were collected only once. Dynamic indicators were constantly varied and collected every morning at 8:00 during the first 14 days of ICU hospitalization. The included variables were listed in **Additional file 1**. Mean ± standard deviation was applied uniformly to express continuous data, and the median (inter-quartile) was for categorical data. Moreover, we conducted a standard procedure of data cleaning, including data alignment, data complementation, and invalid data screening, to make sure that our datasets were of high quality. The details are shown in **Additional file 2.**

*Statistical analysis*

The continuous data was analyzed with t-test if it fitted the normal distribution,

otherwise, the Wilcoxon test (rank sum test) was used. For categorical data, the chi-square test or Wilcoxon test (rank sum test) was used. The correlation analysis was applied to choose the features which exhibit strong correlations with the daily energy target of septic patients. According to the data distribution, we conducted the correlation analysis via the Spearman analysis, and defined strong correlations as when the p value was lower than 0.01. If there is a disagreement, we invited senior physicians (HJ, Wei Chen and LC) to discuss and make decisions. We conducted the statistical analysis by SPSS 26 (IBM, Chicago, USA).

*Modeling*

The metabolic pathophysiological status after sepsis can be divided into three phases, the data-set is divided to three sub-datasets consequently (**Figure 1**). 错误!未找到引用源。

The line graph showed that Trends in mortality with increasing daily energy target which was inputted in the models over early period of acute phase (red), late period of acute phase (orange), and rehabilitation phase (green). The dots showed the lowest point of this curve, which were mapped to the vertical and horizontal axes indicates the lowest mortality and the optimal energy target in the phases, respectively.

These three sub-datasets were called early period of acute phase, late period of acute phase and rehabilitation phase, and are represented data from 1-2, 3-7, and 8-14-days during the entire data collection period. We adopted the Convolutional Neural Networks (CNN) method to build the prognostic models of these three phases. and predicted the optimal energy targets which caused the lowest mortality. The modeling datasets were separated into a training set and a testing set in a 7:3 ratio. A training set was applied to build the model, and the testing set was used for validation by running the models and predicting the optimal energy target of each phase. (The schematic is shown in **Figure 2**). In this study, Python 3.8 software was applied for prediction model building.

RESULTS

*Data collection*

A total of 208 retrospective samples from September 2018 to January 2020 were included in this study as the datasets used to establish the models. To ensure the accuracy of the interpolated data, samples with more than 30% missing data for the same variable were excluded, and a total of 179 patients with sepsis data were finally included in the study. Of these, 78 patients died, and 101 patients survived at least 14-day after admission to the ICU. After establishing the models successfully, we further enrolled 98 patients who were admitted to the ICU from September 2020 to January 2023 as the external validation datasets to predict the optimal energy target (**Figure 3**).

*Data preprocessing*

The missing rates of data among all features were shown in **Additional file 3** during the 14 days after patients were admitted to the ICU. Procalcitonin (PCT) was well over 30% and as high as 82.45% and was excluded which resulted in 34 features being enrolled in our study.

*Energy intake*

We compared the daily energy target between the deceased patients and surviving patients. The results were shown in **Additional file 4** and **Additional file 5**, which showed that the patients of the survival group had higher daily energy target than those in the death group. In addition, he result of cumulative energy debt showed that both groups were at high risks of malnutrition due to increased energy debt during the whole ICU stay (**Figure 4**).

However, the cumulative energy debt of the deceased group increased between the 7th and 14th days of ICU stay, while that decreased in the surviving group, although the difference was not significant (**Additional file 5**).

*Dimension reduction*

As it was shown in **Table 1**, there were 12 features that are strongly related to the daily energy target. However, the leukocyte count (WBC) and neutrophil count (NEUT) have extremely stable and strong correlations during the whole hospitalization of septic patients, we therefore chose the NEUT as the feature to be used for prediction (**Additional file 6**). In addition, body mass index (BMI) was still the main consideration in the development of nutritional program,[14] and the primary diagnosis of patients and the application of mechanical ventilation were both key influencing factors during the implementation of nutritional support therapy. Thus, we included them in the features. Finally, we enrolled a total of 15 features, including age, BMI, NRS2002, diagnosis number, mechanical ventilation therapy, temperature, diastolic blood pressure (DBP), mean arterial pressure (MAP), urine volume, fluid input, creatinine (Cr), aspartate aminotransferase (AST), NEUT, lactate (Lac) and oxygen saturation (SaO2), to build the prognostic prediction models for the optimal energy target.

*Models*

        **Prognostic prediction models**

By the application of CNN methods, we successfully established the prognostic prediction models for the early period of acute phase (358 data), late period of acute phase (889 data) and rehabilitation phase (837 data) of septic patient. In addition, we employed the under-sampling strategy to make sure that the training set was balanced while building the models.

The results validated our hypothesis and we identified the optimal energy target during each phase. The optimal energy target in early period of acute phase (196 data), late period of acute phase (461 data) and rehabilitation phase (374 data) of septic patients was 900kcal/d, 2300kcal/d and 2000kcal/d, which led to a minimum mortality rate of 16%, 28%, and 21%, respectively (**Figure 5**).

The histogram graph compared the optimal daily energy target (left) and the lowest mortality (right) which was derived from models of early period of acute phase (red), late period of acute phase (orange), and rehabilitation phase (green).

As shown in **Figure 6**, mortality raised rapidly while the energy intake fell below the optimal energy target (2300kcal/d) in the late period of the acute phase. In contrast, excessive energy intake increased the mortality rapidly in the early period of acute phase. For the rehabilitation phase, energy targets that are too high or too low both caused high mortality.

**DISCUSSION**

This study demonstrates that nutritional debits are strongly associated with the risk of death during ICU stay of septic patients. In addition, according to the updated guideline critically ill patients by ESPEN, full nutrition supplement are not appropriate in early stage of the illness, but should be gradually achieved within 3-7 days of ICU admission. [15] This study provided strong evidence for the above recommendation and found that excessive energy delivery in the early period and under-energy delivery in the late period of the acute phase could lead to worse prognosis for septic patients. For the rehabilitation phase, both have caused high mortality.

Permissive underfeeding strategy of nutrition support was proposed and discussed for decades.[16] With the growing attention of nutrition support, researchers found that early full-feeding was associated with better clinical outcome of critically ill septic patients.[17] However, subsequent evidence showed that high-energy intake in the acute period is detrimental to the recovery of critically ill patients.[18,19] Previous studies indicated the benefit of using permissive underfeeding in the acute phase of sepsis.[20]. In 2004, Jeejeebhoy KN et al. published a review and indicated that permissive underfeeding may optimize the energy delivery for septic patients.[21] In 2014, Anwar Elias Owais et al. conducted a randomized clinical trial and demonstrated that permissive underfeeding was associated with fewer septic complications ($p = 0.003$).[22] Then in 2021, Sun JK et al. conducted a randomized clinical study (RCT) including 54 septic patients and found that early moderate enteral underfeeding (60% of goal requirements) could improve the intestinal barrier function and reduced inflammatory responses.[23] Notably, Mette M. Berger et al. published a comment according to the 'French-Speaking ICU Nutritional Survey' (FRANS) study conducted in 26 ICUs over 3 months in 2015.[24] In the FFRANS study, full-feeding during the first two days in the ICU led to negative outcomes, and the overfeeding (exceeded the 40kcal/kg) may lead to apparent worse outcomes.[25] This comment emphasized the damage of "too much too early" energy intake in the acute phase of critically illnesses and suggested the feeding dose of 10-20kcal/kg during the first days in ICU. Our study provided objective evidence that permission underfeeding may lead to a positive prognosis for septic patients in the first two days of ICU. However, the growing energy debt caused by permissive underfeeding raise the death risk of patients, which may outweigh the benefits. Tomoaki Yatabe published a review in 2019 and declared that negative energy balancing is beneficial to patients during unstable periods, but may turned to be harmful while it accumulate.[26] Thus, the review suggested to increase the patient's energy supply at appropriate times to cover the energy debt. Our study provided direct evidence for above problem. In the late period of the acute phase, appropriate overfeeding may increase the opportunity to survive and safely settle the energy debt caused by permissive underfeeding for septic patients.

In addition, we provided evidence that it is possible to create practical tools which can minimize nutrition-related mortality. As early as 40 years ago, it was discovered that energy expenditure in critically ill patients might be associated with the regular changes of metabolic phases.[27] In 2019, ESPEN published a guideline on clinical nutrition in the intensive care unit and proposed the three phases of metabolism.[5] In fact, the trend of metabolic indicators varied among individuals, and the importance of them changed in different phases.[28] A study in 2021 showed that the importance of features was dynamic in septic patients. WBC was one of the most important features for the prediction of patient death in the early course of sepsis. Then it gradually worked in predicting survival in the late course. Our study conducted further multivariate analysis for the time-series data by partial least squares discriminant analysis (PLS-DA) methods and came to similar conclusions (Additional file 7). With the rapid growth of artificial intelligence (AI) technology, precision nutrition gained widespread attention.[29] Current reviews emphasized the necessity of personalized nutrition support for critically ill patients and provided several recommendations according to expert experiences.[30-33] However, it was difficult to develop a practical tool due to the lack of direct evidence about its feasibility. In our study, we validated the existence of sweet spots during different metabolic phases, and successfully predicted the optimal energy targets for septic patients, which indicated the possibility of developing personalized nutritional tools for critically ill patients.

Although we successfully established the individual prediction models to optimize the energy delivery for septic patients, this study still had limitations. The limitation of sample size made it difficult to apply in clinical practice immediately. In addition, the extrapolation of our results needs to be further verified because we conducted this study in a single center.

## CONCLUSION

This study establishes time-series prediction models for septic patients admitted to ICU to optimize their energy delivery. The study demonstrates the positive influence of permissive underfeeding strategy in the early period of the acute phase of sepsis while emphasizing the importance of settling the energy debt in the late period of the acute phase. When implemented at ICU, this could serve as an invaluable aide for clinicians.

## DECLARATIONS

### *Ethics approval and consent to participate*

Our study has been reviewed by the Medical Ethics Committee of Sichuan Provincial People's Hospital (No. 190 in 2019) and has been registered in the China Clinical Trial Registration Center (Clinical Registration No.: ChiCTR1900024746. Registered 26 July 2019.). This study is an observational study and will not interfere with the treatment plan of patients, the ethics committee waived informed consent.

### *Consent for publication*

Not applicable.

*Availability of data and materials*

The datasets used and analysed during the current study are available from the corresponding author on reasonable request.

*Competing interests*

The authors declare that they have no competing interests.

*Funding*

This work was Supported by Sichuan Science and Technology Program (No.2021YFH0109 to Mingwei Sun and 2021YFS0378 to Hua Jiang) and National Natural Science Foundation of China (No.72074222 to Wei Chen).

*Authors' contributions*

LW was the lead author of the study. LW conceived and delineated the hypotheses, designed the study, acquired, and analyzed the data, and wrote and edited the manuscript of the previous analysis. HJ, JZ, WC, M-WS and LC developed the original concepts for this study. LW, K-XL, X-LF and R-PZ contributed to the data collection, and data synthesis. R-PZ also contributed to the model building. LW and YW drafted the first version of this manuscript. All authors read and approved the final manuscript and take responsibility for its publication. Dr. Charles Damien Lu and QW contributed to the English proofreading and editing.

*Acknowledgements*

The authors appreciate Weijie Wang for his kind help with the model building.

**Table 1 The results of correlation analysis between the daily energy target and other features.**

| Feature | Coefficient | P value | Feature | Coefficient | P value |
|---|---|---|---|---|---|
| Age | -0.060** | 0.000 | Urea | 0.034* | 0.023 |
| Height | 0.023 | 0.138 | Cr | -0.134** | 0.000 |
| Weight | -0.005 | 0.744 | Glu | 0.017 | 0.258 |
| BMI | -0.021 | 0.165 | Alb | 0.031* | 0.035 |
| SOFA | 0.016 | 0.297 | AST | -0.057** | 0.000 |
| NRS2002 | -0.054** | 0.002 | ALT | 0.01 | 0.508 |
| Nutricscore | -0.013 | 0.431 | TB | -0.007 | 0.634 |
| APACHE II | 0.012 | 0.418 | WBC | -0.052** | 0.001 |
| Temperature | 0.048** | 0.002 | NEUT | -0.060** | 0.000 |
| HR | -0.026 | 0.081 | PLT | 0.022 | 0.135 |
| RR | 0.009 | 0.569 | hs-CRP | -0.019 | 0.213 |
| SBP | -0.011 | 0.443 | Lactate | -0.051** | 0.001 |
| DBP | -0.044** | 0.004 | $S_aO_2$ | 0.061** | 0.000 |
| MAP | -0.033* | 0.026 | | | |
| SI | -0.004 | 0.811 | | | |
| Urine volume | 0.102** | 0.000 | | | |
| Fluid intake | 0.217** | 0.000 | | | |
| Fluid input | 0.019 | 0.191 | | | |

BMI: body mass index; HR: heart rate ; RR: respiratory rate ; SBP: systolic blood pressure; DBP: diastolic blood pressure; MAP: mean arterial pressure; SI: shock index; Cr: creatinine; Glu: blood glucose; AST: aspartate aminotransferase; ALT: alanine aminotransferase; Alb: albumin; TB: total bilirubin; WBC: leukocyte count; NEUT: neutrophil count; PLT: platelet count; hs-CRP: hypersensitive C-reactive protein; Lac: lactate; $S_aO_2$ : oxygen saturation

**: P≤0.001

*：P≤0.05

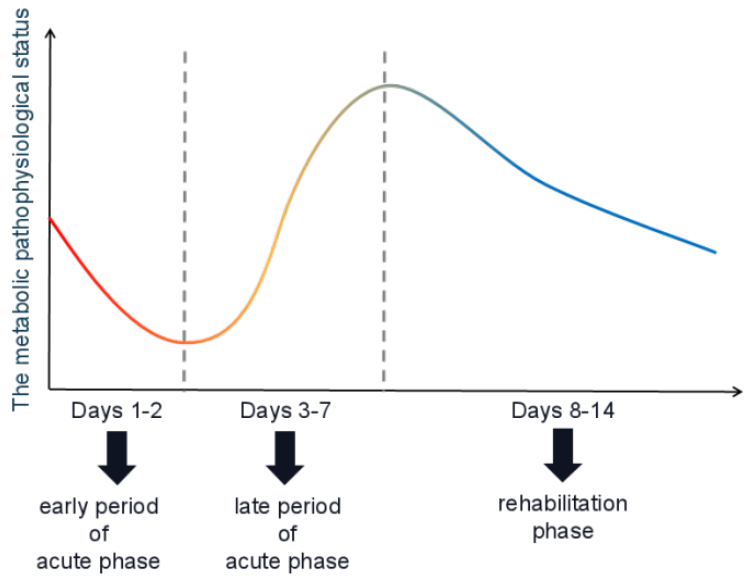

Figure 1    The description of the early period of acute phase, late period of acute phase and rehabilitation phase.

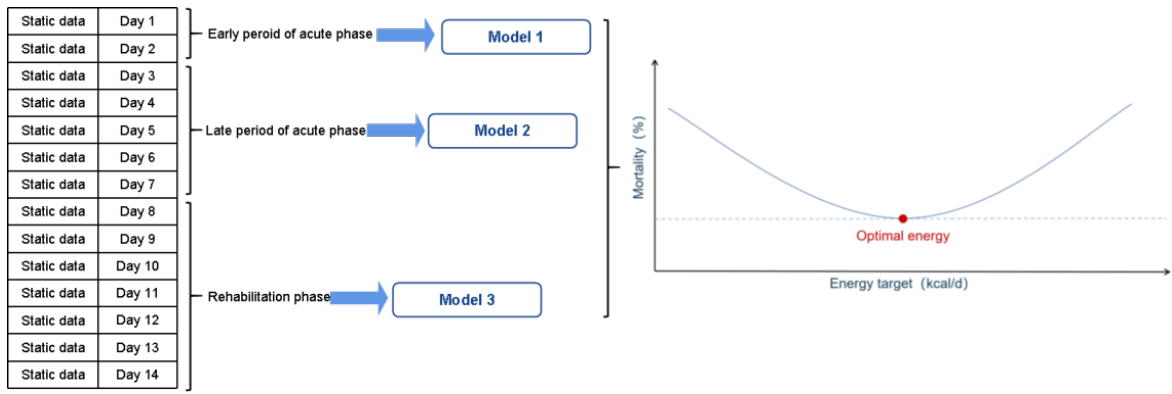

Figure 2    The procedure to build the prediction model of the optimal energy target

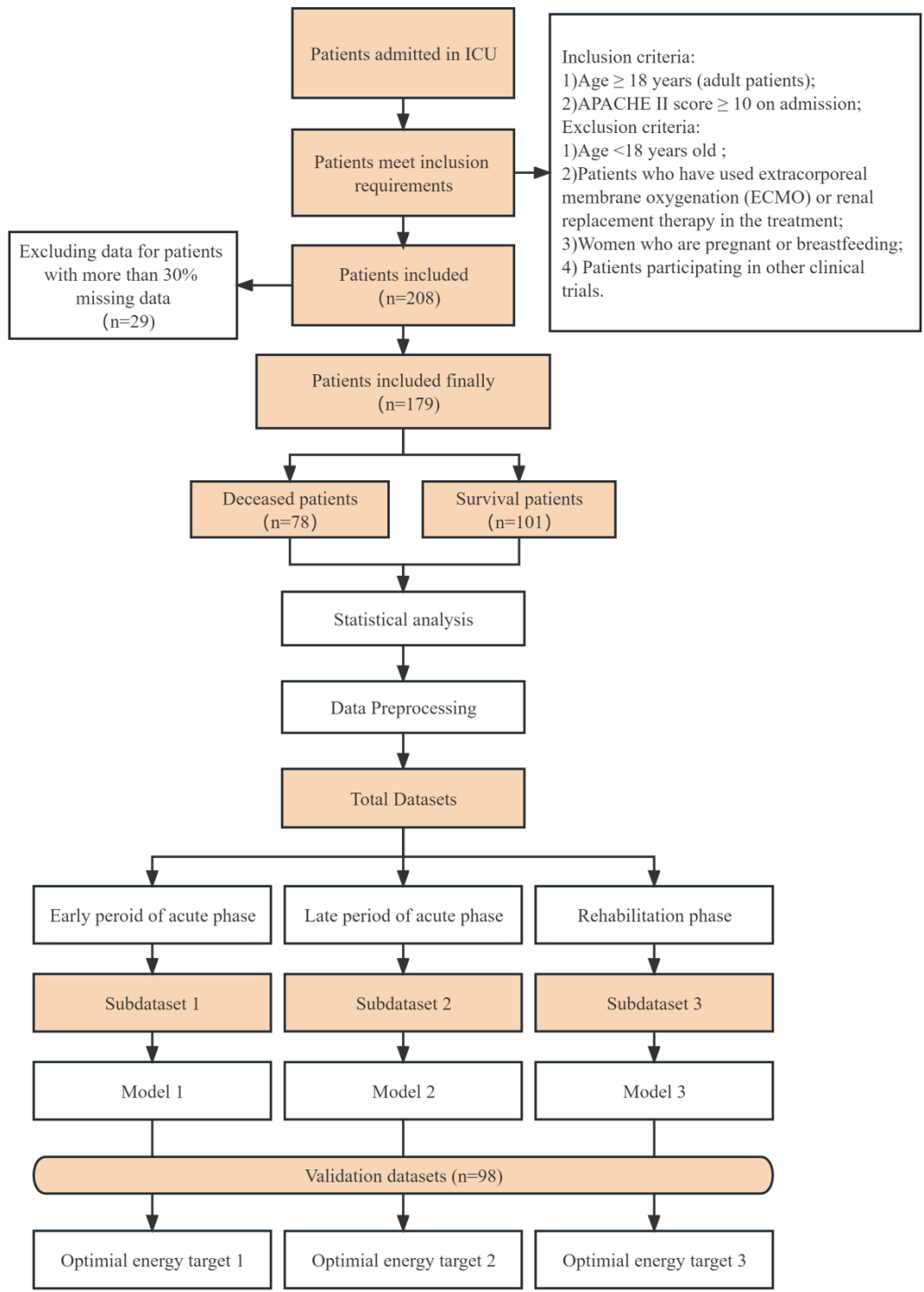

Figure 3　The flowchart of patient enrollment data collation.

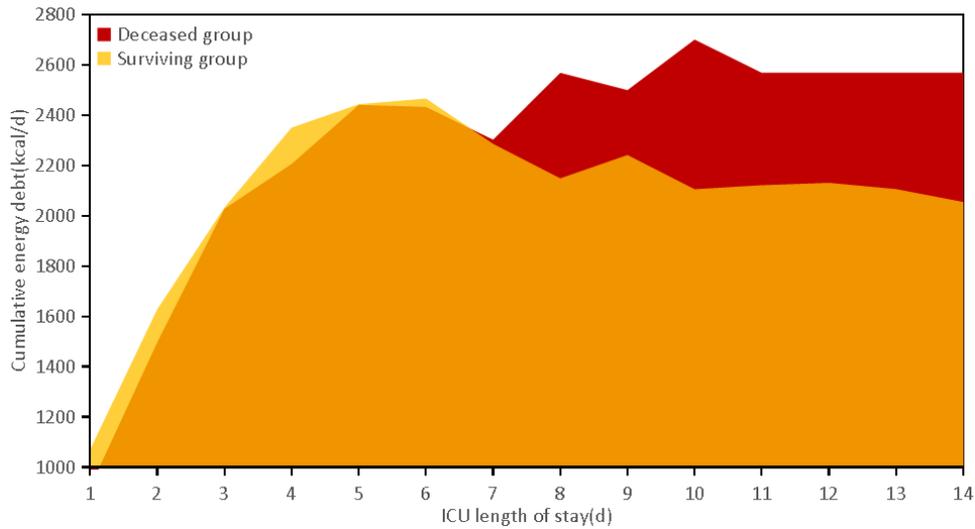

Figure 4　Trend of energy debt over time in deceased and surviving groups.

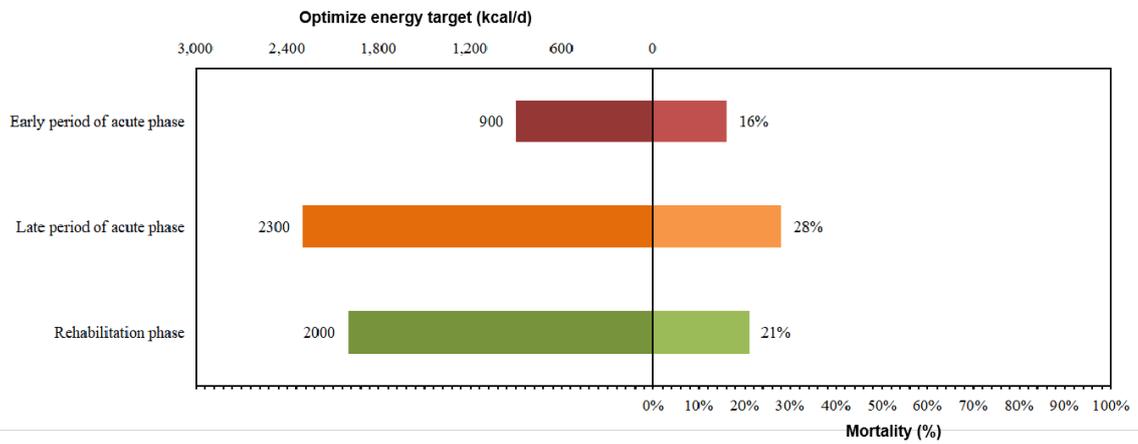

Figure 5　The comparison of the optimal daily energy target and mortality of each phase.

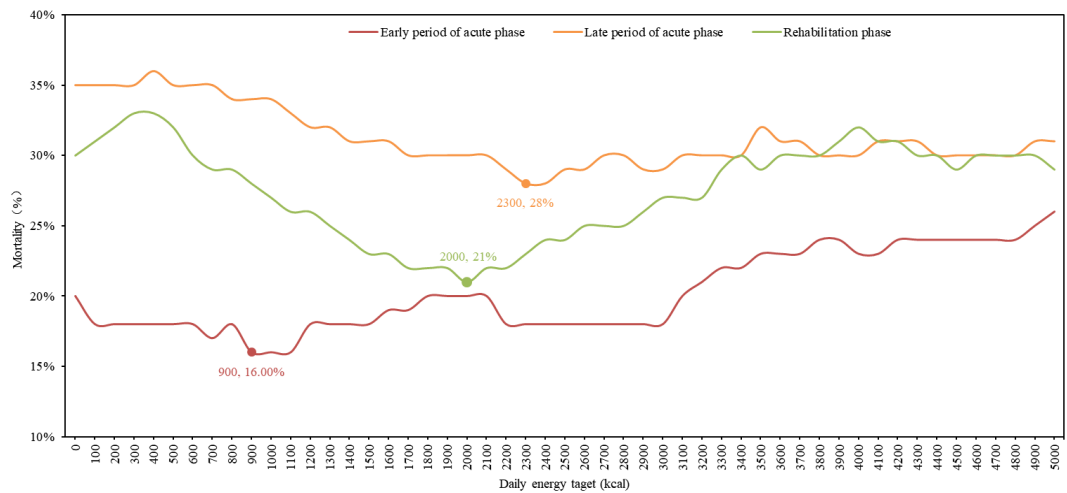

Figure 6　The trends in mortality with increasing daily energy target for septic patients during each phase.

**Text S1.** The variables included in this study.
1) Static variables: including gender, age, primary diagnosis, mental state, mechanical Ventilation, height,weight, body mass index(BMI), Acute Physiology and Chronic Health Evaluation II(APAHCEII), Sepsis-related Organ Failure Score(SOFA). All the static variables were collected only one time at admission or discharge by electronic medical record system.
2) Dynamic variables: including physiological indicator variables, biochemical indicator variables and daily nutritional support plan.
    a. physiological Indicator Variables: temperature, heart rate(HR), respiratory rate(RR)，systolic blood pressure(SBP), diastolic blood pressure(DBP),daily urine volume,daily fluid intake,daily fluid input. The variables created before 2019 were captured through the Electronic Medical Record (EMR), and those created after 2019 are captured through the DoCare critical care system.
    b. biochemical variables: white blood cell count (WBC), neutrophil count (NEUT), hypersensitive C-reactive protein (hs-CRP), platelet count (PLT) , aspartate aminotransferase (AST), alanine aminotransferase (ALT), urea, creatinine (Cr), glucose (Glu), albumin (ALB), total bilirubin (TB), direct bilirubin (DB), indirect bilirubin (IB), lactate (Lac), oxygen saturation ($S_aO_2$). In addition, blood gas analysis indicators, and inflammatory indicators such as procalcitonin (PCT) were all collected. The blood gas analysis results were collected manually, and the rest of the indicators were collected through the Electronic Medical Record (EMR).
    c. nutritional support indicator variables: include nutritional approach and total amount of glucose given by both parenteral and enteral approach.

    In addition, some composite indicators are calculated from the above variables with the following equations.

$$BMI = \frac{Weight（kg）}{Height（m）^2} \quad (2)$$

$$Mean\ Arterial\ Pressure(MAP) = \frac{SBP + DBP \times 2}{3} \quad (3)$$

$$Shock\ Index\ (SI) = \frac{SBP}{HR} \quad (4)$$

$$\frac{收缩压}{心率}$$

Energy(from parenteral or enteral approach)(kcal)=Glucose(g) × 4 + Lipids(g) × 9  (5)

Total Energy(kcal)=Energy from parenteral approach(kcal)+Energy from enteral approach(kcal)  (6)

**Text S2.** The process of data cleaning.

1. **Alignment**

   As we mentioned above, most of our variables were dynamic. Thus, datasets which following time series can greatly help us for discovering the important information of data. We aligned the dynamic data following the ICU-admitted days，and reorganize the data according to the time-series of each patient.

2. **Data Preprocessing**

   As a real world clinical study (RWS), it is difficult to avoid the data missing. However, a high percentage of missing data in any variable can significantly reduce the accuracy of the ML models. We defined the variable (or sample) as invalid data if it met anyone of the following conditions: 1)variables with more than 30% missing data on any given day; 2) samples with more than 30% missing data on any variable within the 14 days after admission in ICU; 3) duplicate data; 4) the data from patients who were not admitted to the ICU for the first time during this hospitalization. All invalid value should be deleted and not participate in any analysis that follows. For the remaining part of missing data, we adopted the mean imputation method for static variables and cubic spline interpolation method for dynamic variables to conduct the data complementation.

3. **Data standardization**

   Different characteristics often have different units of magnitude. For healthcare data, they can vary greatly from one organization to the next and are collected for different purposes. Moreover, these data stored in different formats using different database systems, and the same concept may be represented in different ways from one setting to the next. Data standardization is the critical process of bringing data into a common format that allows for ML methods. In this study, the mean-variance normalization method will be used to standardize the continuous variables, meanwhile, One-Hot Encoding (OHC) is for attribute data which are not well handled by the classifier.

**Table S1. Missing rates of data among all features**

| Feature | Missing rate | Feature | Missing rate |
|---|---|---|---|
| Temperature | 0.06% | Urea | 0.87% |
| HR | 0.06% | Cr | 3.62% |
| RR | 0.06% | Glu | 3.19% |
| SBP | 0.06% | Alb | 3.75% |
| DBP | 0.87% | AST | 3.19% |
| Daily urine volume | 0.81% | ALT | 3.19% |
| Daily fluid intake | 0.81% | TB | 3.19% |
| Daily fluid input | 2.87% | PCT | 82.45% |
| Height | 8.94% | WBC | 3.59% |
| Weight | 24.58% | NEUT | 3.62% |
| BMI | 24.58% | PLT | 3.62% |
| APACHE II score | 4.47% | hs-CRP | 3.62% |
| $S_aO_2$ | 5.60% | Lac | 5.05% |

PCT: procalcitonin; BMI: Body mass index; HR: Heart rate ; RR: Respiratory rate ; SBP: Systolic blood pressure; DBP: Diastolic blood pressure; MAP: Mean arterial pressure; SI: Shock index; Cr: Creatinine; Glu: Blood glucose; AST: Aspartate aminotransferase; ALT: Alanine aminotransferase; Alb: Albumin; TB: Total bilirubin; WBC: Leukocyte count; NEUT: Neutrophil count; PLT: Platelet count; hs-CRP: Hypersensitive C-reactive protein; Lac: Lactate; SaO2 : Oxygen saturation

Figure S1：Trends in daily energy intakes over time for deceased and surviving patients, where blue is the curve for surviving patients and red is the curve for deceased patients. The green vertical dashed line indicates a significant difference (P<0.05) between the energy intake of the two groups of patients at their corresponding days.

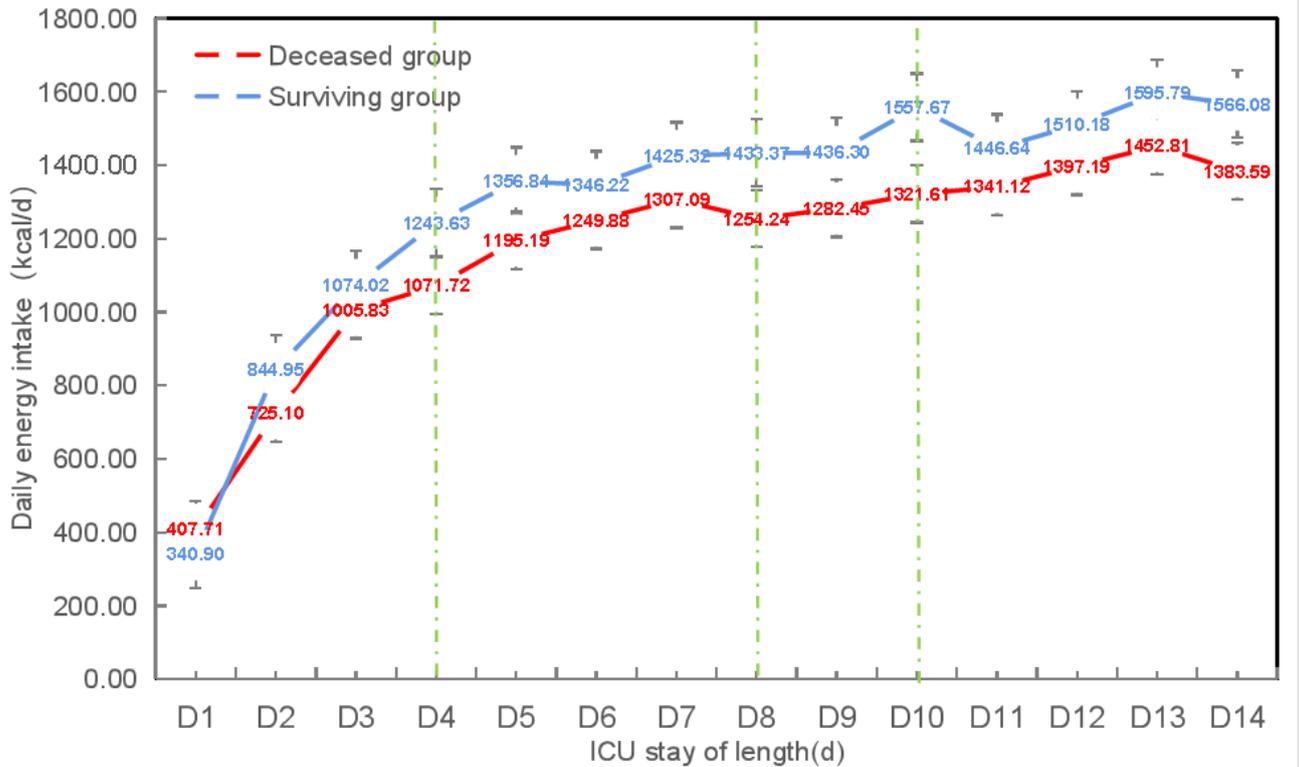

Table S2. Comparison of daily energy target and between deceased and surviving patients

| The length of ICU stay | Total | Deceased group | | Surviving group | | P value |
|---|---|---|---|---|---|---|
| | | No. | Mean ± standard deviation | No. | Mean ± standard deviation | |
| daily energy target (kcal/d) | | | | | | |
| D1 | 179 | 78 | 407.71±61.86 | 10 | 340.90±48.05 | 0.395 |
| D2 | 179 | 78 | 725.10±52.73 | 10 | 844.95±76.08 | 0.224 |
| D3 | 179 | 78 | 1005.83±48.1 | 10 | 1074.02±70.64 | 0.454 |
| D4 | 179 | 78 | 1071.72±50.11 | 10 | 1243.63±70.02 | 0.047 |
| D5 | 179 | 78 | 1195.19±52.03 | 10 | 1356.84±69.97 | 0.065 |
| D6 | 178 | 78 | 1249.88±68.06 | 10 | 1346.22±60.88 | 0.293 |
| D7 | 173 | 75 | 1307.09±73.00 | 98 | 1425.32±64.46 | 0.227 |
| D8 | 161 | 71 | 1254.24±61.53 | 90 | 1433.37±64.52 | 0.046 |
| D9 | 149 | 67 | 1282.45±71.89 | 82 | 1436.30±71.07 | 0.13 |
| D10 | 131 | 60 | 1321.61±73.15 | 71 | 1557.67±71.31 | 0.022 |
| D11 | 122 | 56 | 1341.12±67.94 | 66 | 1446.64±82.99 | 0.327 |
| D12 | 112 | 53 | 1397.19±76.81 | 59 | 1510.18±90.85 | 0.344 |
| D13 | 90 | 40 | 1452.81±83.33 | 50 | 1595.79±80.09 | 0.219 |
| D14 | 81 | 35 | 1383.59±96.54 | 46 | 1566.08±89.67 | 0.17 |
| Cumulative energy debt during ICU admission (kcal) | | | | | | |
| D1 | 179 | 78 | 1299(923,1625) | 10 | 1480(853,1533) | 0.320 |
| D2 | 179 | 78 | 2234(1407,2969) | 10 | 1966(1310,2544) | 0.010 |
| D3 | 179 | 78 | 2893(1613,3809) | 10 | 2188(1547,3307) | 0.000 |
| D4 | 179 | 78 | 3366(2065,4460) | 10 | 2243(1257,4110) | 0.000 |
| D5 | 179 | 78 | 3608(2062,5084) | 10 | 2431(975,4514) | 0.010 |
| D6 | 178 | 78 | 4033(1937,5479) | 10 | 2366(1036,4850) | 0.000 |
| D7 | 173 | 75 | 4350(1347,5640) | 98 | 2300(657,5120) | 0.000 |
| D8 | 161 | 71 | 4296(860,6010) | 90 | 2383(420,5388) | 0.000 |
| D9 | 149 | 67 | 4178(660,6243) | 82 | 2490(272,5240) | 0.010 |
| D10 | 131 | 60 | 4080(867,6456) | 71 | 2736(-38,5448) | 0.010 |
| D11 | 122 | 56 | 4160(839,6999) | 66 | 2747(-276,5914) | 0.050 |
| D12 | 112 | 53 | 4200(924,7388) | 59 | 2599(-476,5957) | 0.320 |
| D13 | 90 | 40 | 4432(585,7820) | 50 | 3018(-676,5971) | 0.300 |
| D14 | 81 | 35 | 3891(62,7655) | 46 | 3337(-876,6033) | 0.510 |

*P value≤0.05

Table S3. The results of correlation analysis.

| | Age | Height | Weight | BMI | APACHE II | SOFA | NRS2002 | Nutricsoore | T？ | HR | RR | SBP | DBP | MAP | SI | Urine volume | Fluid intake | Fluid input | Urea | Cr | Glu | Alb | AST | ALT | TB | WBC | NEUT | PLT | hs-CRP | lactate | SaO2 | Daily energy target |
|---|---|---|---|---|---|---|---|---|---|---|---|---|---|---|---|---|---|---|---|---|---|---|---|---|---|---|---|---|---|---|---|---|
| Age | 1 | -.352** | -.181** | -0.033 | .176** | -.092** | .504** | .515** | -.070** | -.130** | -.052* | .073** | -.212** | -.108** | -.145** | -.236** | -.056* | -.176** | .397** | .369** | .253** | -.131** | -.112** | -.275** | -.207** | -.043* | -0.006 | -.185** | -0.013 | 0.042 | -0.026 | -.087** |
| Height | | 1 | .570** | 0.031 | -.156** | -0.038 | -.227** | -.296** | 0.013 | .066** | .101** | .072** | .140** | .132** | 0.011 | .169** | .048* | .061** | -.093** | -.073** | -.107** | -0.019 | 0.031 | .108** | .087** | -.062** | -.071** | .100** | .068** | -.101** | -.050* | 0.033 |
| Weight | | | 1 | .793** | -.060** | -0.035 | -.190** | -.094** | .055* | 0.019 | .082** | .178** | .131** | .178** | -.093** | .129** | -.049* | .104** | 0 | .110** | 0.042 | -0.031 | .073** | .117** | .122** | -.066** | -.085** | .099** | .055* | -.048* | -.132** | -0.007 |
| BMI | | | | 1 | .060** | -0.005 | -.099** | .088** | .074** | -0.019 | 0.013 | .162** | .070** | .129** | -.117** | .062** | -.079** | .077** | .069** | .202** | .112** | -0.017 | .067** | .073** | .098** | -0.036 | -.058** | .068** | 0.033 | -0.008 | -.124** | -0.028 |
| APACHE II | | | | | 1 | .351** | .413** | .514** | .139** | .078** | -.063** | 0.004 | -0.006 | -0.004 | .063** | .048* | 0.003 | 0.017 | .249** | .241** | .069** | -.057** | .047* | -0.007 | -.072** | 0.01 | 0.012 | -.186** | .048* | .069** | .084** | 0.017 |
| SOFA | | | | | | 1 | .233** | .415** | .064** | 0.006 | -.098** | 0.007 | -0.037 | -0.025 | 0.002 | 0.041 | -.065** | .108** | .199** | .131** | -0.029 | -.059** | .069** | -0.018 | .246** | 0.022 | 0.01 | -.259** | .077** | .119** | .095** | 0.025 |
| NRS2002 | | | | | | | 1 | .414** | 0.026 | -.060** | -.053* | .058** | -.044* | -0.005 | -.081** | -.121** | .092** | -.183** | .203** | .189** | 0.041 | -0.037 | -0.02 | -.117** | -.275** | -0.007 | -0.009 | -.083** | -.050* | -0.036 | .134** | -.069** |
| Nutricsoore | | | | | | | | 1 | -0.013 | -.113** | -.120** | .129** | -.202** | -.068** | -.157** | -.089** | -.150** | 0.056 | .319** | .313** | .182** | -.107** | -.071** | -.217** | -.050* | -.093** | -.071** | -.255** | -0.008 | .181** | 0.023 | -0.018 |
| T？ | | | | | | | | | 1 | .283** | .097** | 0.003 | 0.005 | 0.003 | .216** | .125** | .060** | .104** | .094** | .105** | 0.024 | -.065** | .124** | .128** | .075** | .055* | .052* | -0.037 | .188** | .081** | -.095** | .070** |
| HR | | | | | | | | | | 1 | .286** | -.081** | .154** | .059** | .813** | -0.023 | -0.017 | .105** | .084** | .084** | -0.019 | -.051* | .054* | .067** | .050* | .209** | .209** | 0.008 | .176** | .129** | -.106** | -0.039 |
| RR | | | | | | | | | | | 1 | 0.012 | 0.022 | .210** | -.045* | .086** | -0.021 | -0.025 | -0.039 | -.067** | -0.032 | -0.006 | .101** | -0.01 | .049* | .048* | .114** | 0.029 | 0.003 | -.139** | 0.011 |
| SBP | | | | | | | | | | | | 1 | .410** | .773** | -.608** | .088** | -0.04 | 0.023 | .049* | .088** | .072** | 0.035 | -.059** | -.074** | -0.003 | -.043* | -0.037 | -0.006 | -0.001 | -0.008 | -0.039 | -0.017 |
| DBP | | | | | | | | | | | | | 1 | .878** | -.109** | .055* | .140** | -.148** | -.153** | -.111** | -0.031 | .109** | 0.035 | .097** | 0.04 | 0.041 | 0.022 | .160** | -.065** | -.111** | 0.015 | -.064** |
| MAP | | | | | | | | | | | | | | 1 | -.377** | .083** | .071** | -.081** | -.080** | -0.032 | 0.021 | .092** | -0.005 | 0.029 | 0.021 | 0.001 | -0.01 | -.049** | .075** | -.075** | -0.012 | -.048* |
| SI | | | | | | | | | | | | | | | 1 | -.053* | 0.015 | .079** | 0.042 | 0.007 | -.062** | -.055* | .066** | .089** | .045* | .183** | .180** | 0.007 | .138** | .105** | -.055* | -0.005 |
| Urine volume | | | | | | | | | | | | | | | | 1 | .076** | .324** | -.053* | -.065** | 0.022 | -.061** | 0.039 | .076** | .092** | -.046* | -0.04 | -.087** | .118** | 0.038 | 0.007 | .148** |
| Fluid intake | | | | | | | | | | | | | | | | | 1 | -.487** | .069** | -.117** | -0.007 | .127** | -0.025 | 0.054* | -.202** | -.050* | -.073** | .158** | -.197** | -.207** | .109** | .311** |
| Fluid input | | | | | | | | | | | | | | | | | | 1 | -.063** | -0.003 | 0.002 | -.194** | 0.028 | 0.002 | .292** | .087** | .094** | -.182** | .254** | .275** | -.092** | 0.029 |
| Urea | | | | | | | | | | | | | | | | | | | 1 | .679** | .175** | -.061** | 0.01 | -.080** | -0.012 | 0.039 | .053* | -.266** | .068** | .062** | -0.017 | .051* |
| Cr | | | | | | | | | | | | | | | | | | | | 1 | .153** | -.069** | .046* | -.099** | -.104** | .074** | .075** | -.243** | .199** | .072** | -.077** | -.196** |
| Glu | | | | | | | | | | | | | | | | | | | | | 1 | -0.036 | -0.016 | -.080** | -.064** | -0.033 | -0.002 | -.118** | .047* | .255** | -0.021 | 0.026 |
| Alb | | | | | | | | | | | | | | | | | | | | | | 1 | 0.005 | 0.021 | 0.002 | 0.012 | -0.005 | .224** | -.250** | -.143** | .107** | .046* |
| AST | | | | | | | | | | | | | | | | | | | | | | | 1 | .661** | .261** | .098** | .098** | -.109** | 0.023 | .205** | -0.007 | -.085** |
| ALT | | | | | | | | | | | | | | | | | | | | | | | | 1 | .177** | .095** | .088** | 0.009 | -.044* | .043* | -0.018 | 0.015 |
| TB | | | | | | | | | | | | | | | | | | | | | | | | | 1 | .084** | .079** | -.128** | .176** | .191** | -0.042 | -0.011 |
| WBC | | | | | | | | | | | | | | | | | | | | | | | | | | 1 | .976** | .275** | .098** | .063** | -0.034 | -.077** |
| NEUT | | | | | | | | | | | | | | | | | | | | | | | | | | | 1 | .241** | .120** | .083** | -0.041 | -.090** |
| PLT | | | | | | | | | | | | | | | | | | | | | | | | | | | | 1 | -.089** | -.216** | 0.032 | 0.034 |
| hs-CRP | | | | | | | | | | | | | | | | | | | | | | | | | | | | | 1 | .058** | -.147** | -0.026 |
| lactate | | | | | | | | | | | | | | | | | | | | | | | | | | | | | | 1 | -.058** | -.075** |
| SaO2 | | | | | | | | | | | | | | | | | | | | | | | | | | | | | | | 1 | .088** |
| Daily energy target | | | | | | | | | | | | | | | | | | | | | | | | | | | | | | | | 1 |

*P value＜0.05

**P value＜0.01

Figure S3: The VIP values over time of the basic characteristics whose value > 1: A. The characteristics of a decreasing trend of VIP value over time; B. The characteristics of an increasing trend of VIP value over time; C. Trend of change in VIP values of WBC and platelet count from 5 to 14 days after ICU admission.

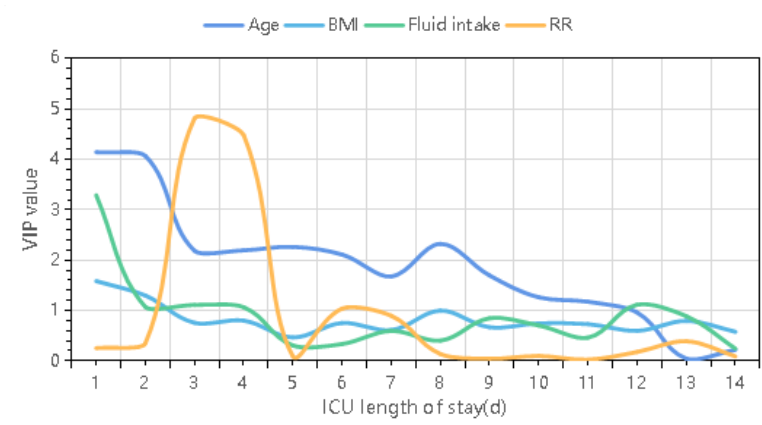

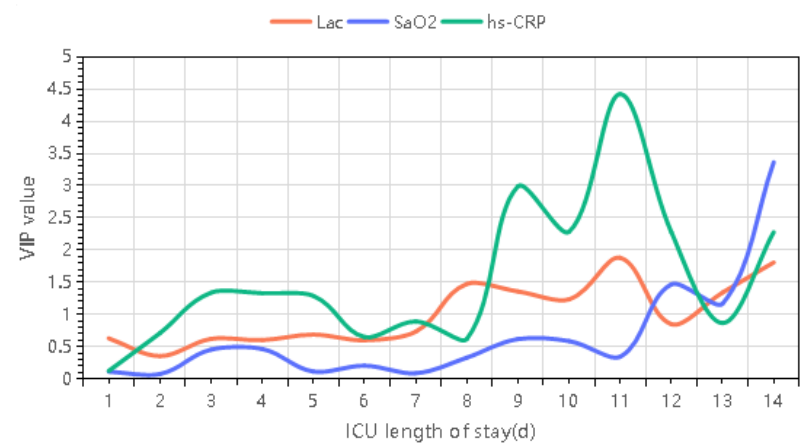

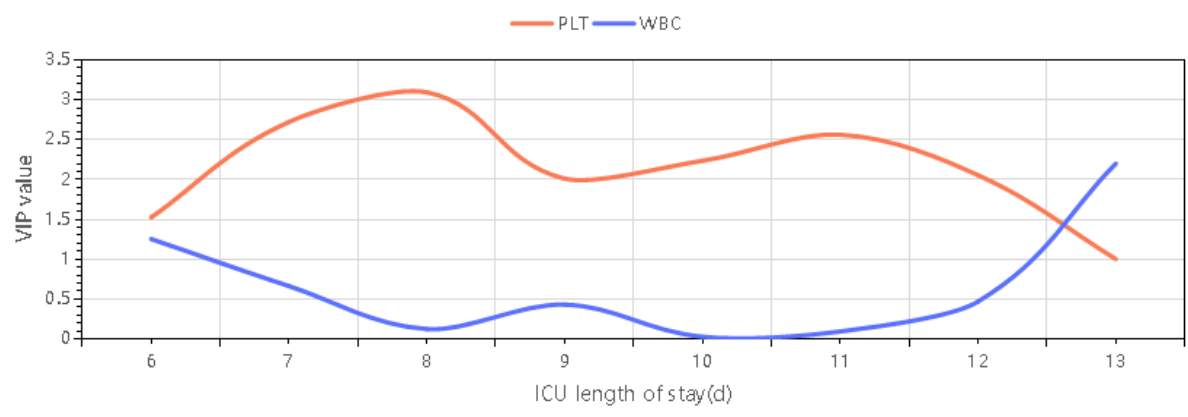